\documentclass[prb,preprint]{revtex4}%
\usepackage{mathrsfs}
\usepackage{subfig}
\usepackage{dsfont}
\usepackage{amsfonts}
\usepackage{amsmath}
\usepackage{amssymb}
\usepackage{graphicx}
\usepackage{epstopdf}
\usepackage{color}%
\begin{document}
	\title{Revisiting the Bragg reflector to illustrate some modern developments in optics}
	\author{S. A. R. Horsley}
	\affiliation{Electromagnetic and Acoustic Materials Group, Department of Physics and Astronomy, University of Exeter, Stocker Road, Exeter, EX4 4QL, UK}
	\email{s.horsley@exeter.ac.uk}
	\author{J.-H. Wu}
	\affiliation{College of Physics, Jilin University, Changchun 130012, P. R. China}
	\author{M. Artoni}
	\affiliation{European Laboratory for Nonlinear Spectroscopy, Sesto Fiorentino, Italy}	
	\affiliation{Department of Engineering and Information Technology CNR-IDASC Sensor Lab, Brescia University, Brescia, Italy}
	\author{G. C. La Rocca}
	\affiliation{Scuola Normale Superiore and CNISM, Pisa, Italy}
	\date{\today}
%
%
\begin{abstract}
	A series of thin layers of alternating refractive index is known to make a good optical mirror over certain bands of frequency.  Such a device---often termed the \emph{Bragg reflector}---is usually introduced to students within the first years of an undergraduate degree, often in isolation from other parts of the course.   Here we show that the basic physics of wave propagation through a stratified medium can be used to illustrate some more modern developments in optics as well as quantum physics; from transfer matrix techniques, to the optical properties of cold trapped atoms, optomechanical cooling, and a simple example of a system exhibiting an appreciable level of optical non--reciprocity.
\end{abstract}

\maketitle

%
%
\section{Introduction}
\par
	To paraphrase Griffiths~\cite{griffiths2001}, wave motion can be divided into at least three kinds.  These are travelling waves (continuous spectrum); bound states (discrete spectrum); and waves within a periodic medium (continuous spectrum with forbidden regions). There are a variety of ways in which the third of this number can be introduced.  Perhaps the most obvious route is through condensed matter physics, where the atomic lattice provides a ubiquitous example of a periodic medium (see e.g.~\cite{ziman1972,volume9,olsen2010}).  To all intents and purposes such a system has an infinite number of periods, and as a consequence Bloch's theorem approaches exactitude.
\par
 	Yet it is instructive to observe the emergence of Bloch's theorem as the number of periods of the system is increased.  Indeed, the motion of waves in a one dimensional periodic, layered medium can be solved exactly for any number of layers~\cite{born1999,griffiths1992,griffiths2001,guo2006}.  A neat example where the onset of a band gap may be observed can be found in the optics of the Bragg reflector~\cite{born1999}: an optical medium composed of layers of two kinds of material with differing refractive indices.  Possibly the most striking lesson one learns through studying an arbitrary number of layers is how quickly the results of Bloch's theorem become applicable as the number of layers is increased from unity.  However, this is not usually the motivation for introducing the Bragg reflector to students; it is often introduced as an interesting optical component within an optics course, and more completely only within a specialized course covering photonic crystals or microcavities. 
\par
	Not only is the Bragg reflector useful for discussing the emergence of Bloch's theorem, but it also unveils some subtleties of propagation in a periodic medium.  Dispersion and dissipation significantly alter the picture, and dissipation in particular can cause a breakdown of the distinction between allowed and forbidden bands of frequency.  Here we briefly derive the optical properties of the Bragg reflector through an application of the transfer matrix formalism~\cite{soto2012}, and the results are applied to the discussion of the optical properties of a lattice of cold trapped atoms~\cite{bloch2005}, where both dispersion and dissipation are significant at the frequencies of interest.  As we shall see, the picture becomes quite different when these effects are included.
	\par
	Further to this we show that the Bragg reflector can be used to introduce the principles of two phenomena of current interest: optomechanical cooling~\cite{marquardt2009}; and optical non--reciprocity~\cite{potton2004,dai2012}.  In both cases the high frequency sensitivity close to a band edge significantly couples the optical response to the centre of mass motion of the reflector, and thereby provides a simple example system where such effects can be exactly described.
	
%
%
\section{Stratified media and transfer matrices\label{tm-section}}
	\par
	To solve the wave equation in a medium that is inhomogeneous in one direction only, one reduce the problem to that of multiplying together \(2\times 2\) matrices.  This is known as the \emph{transfer matrix} technique~\cite{born1999,soto2012}.  In the case of electromagnetic waves in periodic, layered media the formalism is particularly neat, and provides a straightforward means to exactly describe the optics of a Bragg reflector.  The technique is introduced here only for completeness, and the discussion will be brief.  For a more in--depth exposition see the recent paper by S\'anchez--Soto et. al.~\cite{soto2012}.
	\par
\begin{figure}[h!]
	\begin{center}
	\includegraphics[width=8cm]{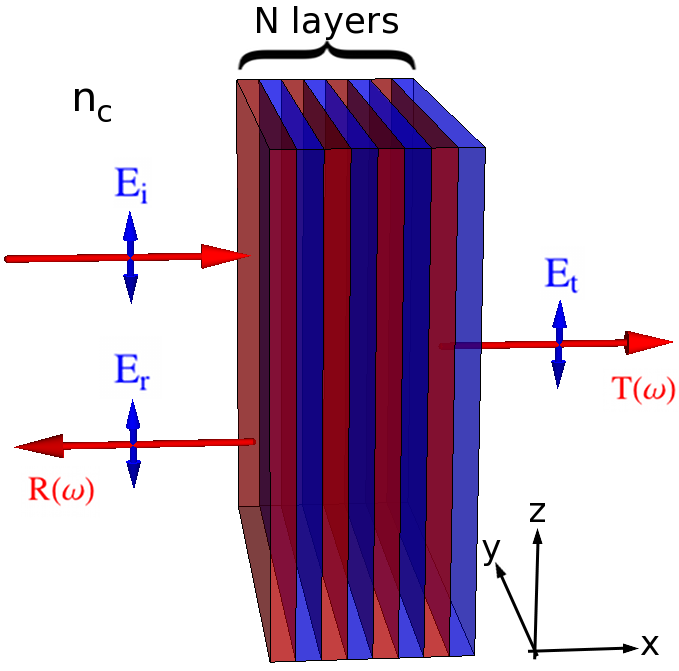}
	\end{center}
	\caption{
	An \(N\) layer structure is surrounded by a medium with a real refractive index \(n_{c}\).  In the rest frame of the medium, we imagine linearly polarized light of frequency, \(\omega\) is incident from the left with electric field amplitude, \(\boldsymbol{E}_{i}\), some of which is reflected as \(\boldsymbol{E}_{r}\) (reflectivity, \(R(\omega)\)), and some of which is transmitted as \(\boldsymbol{E}_{t}\) (transmittivity, \(T(\omega)\)).\label{figure_1}}
\end{figure}

	The system under consideration is illustrated in figure~\ref{figure_1}, where we have a medium that is composed of \(N\) layers, each with a different value of the refractive index, \(n\), immersed in background medium (perhaps a fluid) with index \(n_{c}\).  In general the refractive indices of the layers will be complex numbers, the imaginary parts of which indicate the degree to which the electromagnetic field is absorbed.  For simplicity we assume normal incidence, where propagation through the system does not depends on the polarization of the waves.
%
%
	\par
	For a fixed frequency of oscillation, \(\omega\) the solutions to the wave equation within each layer of the medium are plane waves, with spatial dependence \(\exp{(\pm i k_{xm}x)}\) (\(m\) labels the layer).   The electric field, \(\boldsymbol{E}\) at every point in space is then written as a sum of two parts; a \emph{right} moving wave, \(e^{(+)}_{m}(x)=A^{(+)}_{m}e^{ik_{xm}x}\), and a \emph{left} moving wave, \(e^{(-)}_{m}(x)=A^{(-)}_{m}e^{-ik_{xm}x}\), with the understanding that the full field is the sum of the two,
	\begin{equation}
		\boldsymbol{E}(x)=\hat{\boldsymbol{p}}\left[A^{(+)}_{m}e^{ik_{xm}x}+A^{(-)}_{m}e^{-ik_{xm}x}\right],	
	\end{equation}
	where \(\hat{\boldsymbol{p}}\) is a unit vector denoting the polarization of the waves (here this is the same throughout the medium), and \(x\) is a position within the \(m\)th layer.  If we introduces a set of column vectors, \(\boldsymbol{e}_{m}\), the elements of which are \(e^{(+)}_{m}\) and \(e^{(-)}_{m}\), respectively, then a \(2\times2\) matrix, \(\boldsymbol{M}\) can be found to relate the left and right moving parts of \(\boldsymbol{E}\) on either side of the medium, \(\boldsymbol{e}_{N+1}=\boldsymbol{M}\boldsymbol{\cdot}\boldsymbol{e}_{0}\).  The matrix, \(\boldsymbol{M}\) is called the \emph{transfer matrix}.
	\par
	There are only two kinds of things that happen to the light during propagation through a layered medium: it encounters boundaries, and moves through homogeneous regions.  \(\boldsymbol{M}\) can be thus broken into a product of two types of component transfer matrices, an ``\emph{interface}'' matrix, \(\boldsymbol{W}\), giving the relation between the electric field on each side of a boundary, and a ``\emph{translation}'' matrix, \(\boldsymbol{X}\), giving the relationship between the electric field on each side of a homogeneous region.  In both cases superscripts are introduced into the notation to indicate the layer being referred to.
		\par
		The matrix relating the fields on each side of an interface, \(\boldsymbol{W}\) can be found through applying the condition of continuity of the relevant components of the electromagnetic field across the boundary~\cite{volume8}.  For normal incidence and non--magnetic media, the situation is the same as in a scalar wave theory, where the electric field and its first derivative must be continuous across the boundary.  Applying these two conditions across an interface between index \(n_{a}\) and \(n_{b}\), we have
		\begin{equation}
			e^{(+)}_{a}+e^{(-)}_{a}=e^{(+)}_{b}+e^{(-)}_{b},\label{continuity_1}
	\end{equation}
	and,
	\begin{equation}
		n_{a}\left[e^{(+)}_{a}-e^{(-)}_{a}\right]=n_{b}\left[e^{(+)}_{b}-e^{(-)}_{b}\right].\label{continuity_2}
	\end{equation}
	where we have used the relation, \(\partial e^{(\pm)}/\partial x=\pm i k_{x}e^{(\pm)}\), and \(k_{x}=n\omega/c\), to obtain (\ref{continuity_2}).  Writing (\ref{continuity_1}--\ref{continuity_2}) in matrix form and solving for the vector \(e_{b}^{(\pm)}\) one obtains,
	\begin{equation}
		\left(\begin{matrix}e^{(+)}_{b}\\e^{(-)}_{b}\end{matrix}\right)=\frac{1}{2}\left(\begin{matrix}1+n_{a}/n_{b}&1-n_{a}/n_{b}\\1-n_{a}/n_{b}&1+n_{a}/n_{b}\end{matrix}\right)\left(\begin{matrix}e^{(+)}_{a}\\e^{(-)}_{a}\end{matrix}\right)
\equiv \boldsymbol{W}^{(ab)}\left(\begin{matrix}e^{(+)}_{a}\\e^{(-)}_{a}\end{matrix}\right).
\label{equation_2}	
	\end{equation}
	The matrix to the right of the first equality in (\ref{equation_2}) defines the general relation between the electric field across a boundary between two media.
	\par
 	 The matrix, \(\boldsymbol{X}\), relating the values of the field at \(x_{1}\) and \(x_{2}\) in a homogeneous medium simply multiplies \(e^{(+)}\) and \(e^{(-)}\) by a phase factor, \(\exp{(\pm i k_{x} (x_{2}-x_{1}))}\), e.g. for a medium of index \(n_{a}\),
	\begin{equation}
		\left(\begin{matrix}e_{a}^{(+)}(x_{2})\\e_{a}^{(-)}(x_{2})\end{matrix}\right)=\left(\begin{matrix}e^{ik_{a}(x_{2}-x_{1})}&0\\0&e^{-ik_{a}(x_{2}-x_{1})}\end{matrix}\right)\left(\begin{matrix}e_{a}^{(+)}(x_{1})\\e_{a}^{(-)}(x_{1})\end{matrix}\right)\equiv \boldsymbol{X}^{(a)}(x_{2}-x_{1})\left(\begin{matrix}e_{a}^{(+)}(x_{1})\\e_{a}^{(-)}(x_{1})\end{matrix}\right).\label{equation_5}		
	\end{equation}
	\par
	The matrices within (\ref{equation_2}) and (\ref{equation_5}) can be used to study light propagation in \textit{generic} stratified media, including continuously inhomogeneous media~\cite{born1999}.
%
%
\section{The Bragg reflector\label{bragg-section}}
	\par
	The Bragg reflector (sometimes referred to as a 1D photonic crystal) is a stratified medium composed of a \emph{periodic} sequence of identical \emph{unit cells}~\cite{heavens1960,yeh2005}.  In this case a unit cell is taken as two layers, each having different complex refractive indices, \(n_{a}\) and \(n_{b}\), and thicknesses, \(a\) and \(b\).
	\par
	Say the unit cell of our Bragg reflector starts and finishes within the medium of index \(n_{a}\).  From the discussion of the previous section, the left and right moving parts of the field on either side of this unit cell are related by a matrix, \({\boldsymbol{{\cal M}}}_{1}\), equal to the following product,
	\begin{align}
		{\boldsymbol{{\cal M}}}_{1}&=\boldsymbol{W}^{(ba)}\boldsymbol{\cdot}\boldsymbol{X}^{(b)}(b)\boldsymbol{\cdot}\boldsymbol{W}^{(ab)}\boldsymbol{\cdot}\boldsymbol{X}^{(a)}(a)\nonumber\\
		&=\left(\begin{matrix}
			\left[\cos{(k_{b}b)}+i\alpha_{ab}\sin{(k_{b}b)}\right]e^{ik_{a}a}&-i\beta_{ab}\sin{(k_{b}b)}e^{-ik_{a}a}\\
			i\beta_{ab}\sin{(k_{b}b)}e^{ik_{a}a}&\left[\cos{(k_{b}b)}-i\alpha_{ab}\sin{(k_{b}b)}\right]e^{-ik_{a}a}
			\end{matrix}\right),\label{unit-cell-matrix}
	\end{align}
	where, \(k_{a,b}=n_{a,b}\,\omega/c\), \(\alpha_{ab}=(n_{a}^{2}+n_{b}^{2})/(2n_{a}n_{b})\), and \(\beta_{ab}=(n_{a}^{2}-n_{b}^{2})/(2n_{a}n_{b})\).
	\par
	The transfer matrix associated with \(N\) identical unit cells is equal to the \(N\)th power of (\ref{unit-cell-matrix}), \(\boldsymbol{\cal{M}}_{N}=\boldsymbol{\cal{M}}_{1}^{N}\).  Given that \(\boldsymbol{\cal M}_{1}\) is a \(2\times2\) matrix, it turns out that we can write its \(N\)th power in an elegant form involving Chebyshev polynomials of the second kind~\cite{born1999,griffiths1992,griffiths2001}.  To see this, consider the square of a general \(2\times2\) matrix, \(\boldsymbol{A}\),
	\begin{align}
		\boldsymbol{A}^{2}&=\left(\begin{matrix}a_{11}&a_{12}\\a_{21}&a_{22}\end{matrix}\right)\left(\begin{matrix}a_{11}&a_{12}\\a_{21}&a_{22}\end{matrix}\right)=\left(\begin{matrix}a_{11}^{2}+a_{12}a_{21}&a_{12}(a_{11}+a_{22})\\a_{21}(a_{11}+a_{22})&a_{22}^{2}+a_{12}a_{21}\end{matrix}\right)\nonumber\\[5pt]
			&=\text{Tr}[\boldsymbol{A}]\boldsymbol{A}-\text{det}[\boldsymbol{A}]\boldsymbol{\mathds{1}}_{2}\label{matrix-identity}
	\end{align}
	where to obtain the second line we used, \(a_{11}^{2}+a_{21}a_{12}=a_{11}(a_{11}+a_{22})-(a_{11}a_{22}-a_{12}a_{21})\).  It must thus be possible to write the \(N\)th power of \(\boldsymbol{A}\) in the form, \(\boldsymbol{A}^{N}=a_{N}\boldsymbol{A}+b_{N}\boldsymbol{\mathds{1}}_{2}\): i.e. we can recursively apply (\ref{matrix-identity}) to any power of \(\boldsymbol{A}\), reducing it to this form.  Equating the expressions \(\boldsymbol{A}^{N+1}=a_{N+1}\boldsymbol{A}+b_{N+1}\boldsymbol{\mathds{1}}_{2}\), and \(\boldsymbol{A}^{N+1}=a_{N}\boldsymbol{A}^{2}+b_{N}\boldsymbol{A}\), an application of (\ref{matrix-identity}) shows that the coefficients \(a_{N}\) and \(b_{N}\) satisfy the recursion relations,
	\begin{align*}
		a_{N+1}&=\text{Tr}[\boldsymbol{A}]a_{N}-\text{det}[\boldsymbol{A}]a_{N-1}\\
		b_{N+1}&=-a_{N}\text{det}[\boldsymbol{A}]
	\end{align*}
	Meanwhile, the Chebyshev polynomials of the first kind satisfy~\cite{gradshteyn2000},
	\begin{equation}
		U_{N+1}(z)=2z U_{N}(z)-U_{N-1}(z)\label{chebyshev-recurrence}
	\end{equation}
	In the case where \(\text{det}[\boldsymbol{A}]=1\),  the coefficients, \(a_{N}\) satisfy (\ref{chebyshev-recurrence}), and therefore \(\boldsymbol{A}^{N}=U_{N-1}(z)\boldsymbol{A}-U_{N-2}(z)\boldsymbol{\mathds{1}}_{2}\), where \(z=\text{Tr}[\boldsymbol{A}]/2\).
	\par
	Assuming \(\text{det}[\boldsymbol{\mathcal{M}}_{1}]=1\), which for our purposes is always the case (see e.g. equation (\ref{det-trans})), the \(N\)--cell matrix is thus,
	\begin{equation}
		\boldsymbol{\cal{M}}_{N}=\boldsymbol{\cal{M}}_{1}^{N}=U_{N-1}(z)\boldsymbol{\cal{M}}_{1}-U_{N-2}(z)\boldsymbol{\mathds{1}}_{2},\label{multilayer-result}	
	\end{equation}
	where,
	\begin{equation}
	z=\frac{1}{2}\text{Tr}\left(\boldsymbol{\cal{M}}_{1}\right)=\cos{(k_{b}b)}\cos{(k_{a}a)}-\alpha_{ba}\sin{(k_{b}b)}\sin{(k_{a}a)}.\label{equation_9}	
	\end{equation}
	It is usually practically convenient to compute these Chebyshev polynomials from their relation to the trigonometric functions~\cite{gradshteyn2000}: \(U_{N}(z)=\sin{[(N+1)\arccos(z)]}/\sqrt{1-z^{2}}\).
	\par
	The final expression for the complete transfer matrix, \(\boldsymbol{M}\), for figure~\ref{figure_1} is \(\boldsymbol{\cal{M}}_{N}\) sandwiched between two matrices associated with light passing from the surrounding medium (index, \(n_{c}\)), into the reflector and then out again,
\begin{equation}
		\boldsymbol{M}=\left(\begin{matrix}m_{11}&m_{12}\\m_{21}&m_{22}\end{matrix}\right)=\boldsymbol{W}^{(ac)}\boldsymbol{\cdot}\boldsymbol{\cal{M}}_{N}\boldsymbol{\cdot}\boldsymbol{W}^{(ca)},\label{full-transfer}	
\end{equation}
	The elements of which can be found from the results (\ref{equation_2}), (\ref{unit-cell-matrix}) and (\ref{multilayer-result}),
	\begin{align}
		m_{11}&=U_{N-1}(z)\left[z+i\alpha_{ac}\xi-i\beta_{ac}\beta_{ab}\cos{(k_{a}a)}\sin{(k_{b}b)}\right]-U_{N-2}(z)\nonumber\\
		m_{12}&=-U_{N-1}(z)\left[\beta_{ab}\sin{(k_{b}b)}\left(\sin{(k_{a}a)}+i\alpha_{ac}\cos{(k_{a}a)}\right)-i\beta_{ac}\xi\right]\nonumber\\
		m_{21}&=-U_{N-1}(z)\left[\beta_{ab}\sin{(k_{b}b)}\left(\sin{(k_{a}a)}-i\alpha_{ac}\cos{(k_{a}a)}\right)+i\beta_{ac}\xi\right]\nonumber\\
		m_{22}&=U_{N-1}(z)\left[z-i\alpha_{ac}\xi+i\beta_{ac}\beta_{ab}\cos{(k_{a}a)}\sin{(k_{b}b)}\right]-U_{N-2}(z)\label{transfer-matrix}
	\end{align}  
	where,
	\begin{equation}
		\xi=\sin{(k_{a}a)}\cos{(k_{b}b)}+\alpha_{ba}\cos{(k_{a}a)}\sin{(k_{b}b)},
		\label{equation_18}	
	\end{equation}
	\par
 	The results (\ref{transfer-matrix}) can then be combined to give the desired reflection and transmission coefficients.  For example when light of amplitude \(E_{0}\) is incident onto the medium from the left, the transfer matrix equation reduces to,
	\[
		\left(\begin{matrix}E_{0}t_{+}\\0\end{matrix}\right)=\left(\begin{matrix}m_{11}&m_{12}\\m_{21}&m_{22}\end{matrix}\right)\left(\begin{matrix}E_{0}\\E_{0}r_{+}\end{matrix}\right)
	\]
	and one can solve this for \(r_{+}\) and \(t_{+}\) in terms of the elements of \(\boldsymbol{M}\).  Performing such a calculation, one finds the transmissivity to be equal for incidence from the left and right,
	\begin{equation}
		T_{+}=\left|\frac{1}{m_{22}}\right|^{2}=\left|\frac{1}{U_{N-1}(z)\left[z-i\alpha_{ac}\xi+i\beta_{ac}\beta_{ab}\cos{(k_{a}a)}\sin{(k_{b}b)}\right]-U_{N-2}(z)}\right|^{2}=T_{-}\label{transmission-result}
	\end{equation}
	and the reflectivities,
	\begin{align}
		R_{+}&=\left|\frac{m_{21}}{m_{22}}\right|^{2}=\left|\frac{U_{N-1}(z)\left[\beta_{ab}\sin{(k_{b}b)}\left(\sin{(k_{a}a)}-i\alpha_{ac}\cos{(k_{a}a)}\right)+i\beta_{ac}\xi\right]}{U_{N-1}(z)\left[z-i\alpha_{ac}\xi+i\beta_{ac}\beta_{ab}\cos{(k_{a}a)}\sin{(k_{b}b)}\right]-U_{N-2}(z)}\right|^{2}\nonumber\\
		R_{-}&=\left|\frac{m_{12}}{m_{22}}\right|^{2}=\left|\frac{U_{N-1}(z)\left[\beta_{ab}\sin{(k_{b}b)}\left(\sin{(k_{a}a)}+i\alpha_{ac}\cos{(k_{a}a)}\right)-i\beta_{ac}\xi\right]}{U_{N-1}(z)\left[z-i\alpha_{ac}\xi+i\beta_{ac}\beta_{ab}\cos{(k_{a}a)}\sin{(k_{b}b)}\right]-U_{N-2}(z)}\right|^{2}\label{reflection-result}
	\end{align}
	which are not equal unless the medium is lossless, in which case it is evident from (\ref{transfer-matrix}) that \(m_{12}=m_{21}^{\star}\).  In general \(R_{+}\) and \(R_{-}\) should not be expected to be equal.  One example would be a single unit cell where we have a highly reflective surface backed by a thick layer of strongly absorbing material: incidence onto the front surface would yield a reflectivity close to unity, while incidence onto the reverse would yield a much lower value.  However, in the case of a Bragg reflector with many thin layers and small dissipation, this difference is negligible, and we shall often assume that \(R_{+}\sim R_{-}\) in what follows.  The expressions for \(R_{+}\) and \(T_{+}\) are plotted in figure~\ref{figure-4}, for a Bragg reflector of \(10\) unit cells.
%
%
\subsection{Bloch's theorem in optics\label{bloch}}
	\par
\begin{figure}[h!]
		\begin{center}
\includegraphics[width=7 cm]{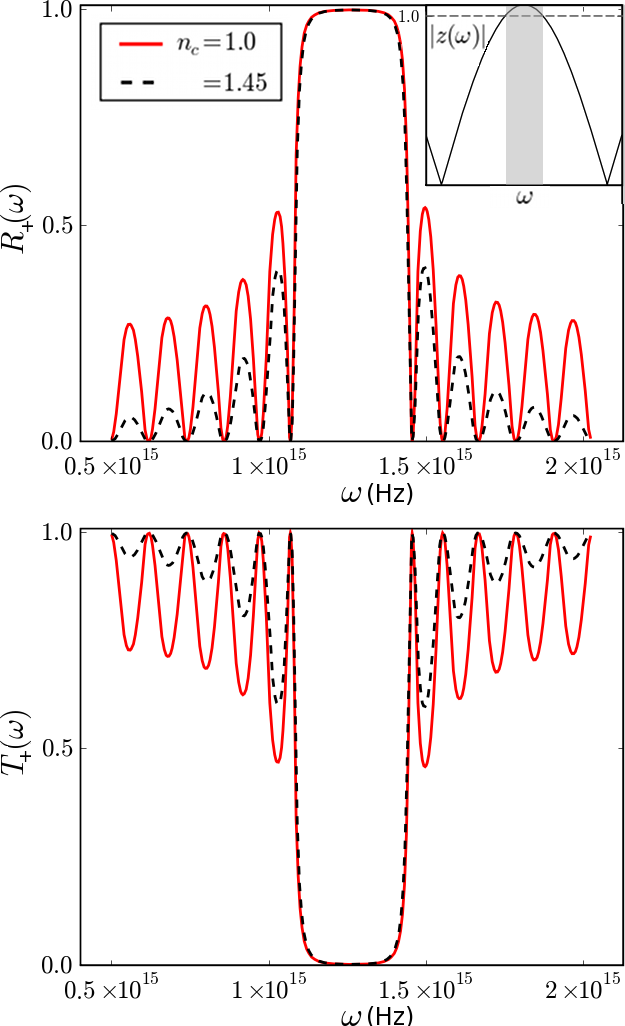}
			\caption{Transmissivity $T_{+}(\omega)$ and reflectivity $R_{+}(\omega)$ as a function of frequency for a \textit{transparent} Bragg reflector, composed of 10 unit cells where all layers have real refractive indices, \(n_{a}=1.45\) and \(n_{b}=2.1\).  The thicknesses are \(a=2.60\times10^{-7}\,\text{m}\) and \(b=1.76\times10^{-7}\,\text{m}\).  The spectra contain a stop-band region where propagation is forbidden, corresponding to the plateau in the central region of \(R_{+}\), or to the grey shaded region where \(\left|z\right|>1\) (inset).  The amplitude of the oscillations increases with the index contrast from \(n_c=1 \to n_c=1.45\) at the edge of the medium, but this does not affect the stop band structure.  Such an increase in oscillation amplitude can be important (see section~\ref{example-section}).  \emph{Inset}: For an infinite array of unit cells no mode can propagate in the frequency range shaded in grey (\(|z|>1\), corresponding to the photonic band gap.\label{figure-4}}
		\end{center}
	\end{figure}

	Any real multilayer system will have a finite number of layers; a Bragg reflector may only have tens or hundreds.  It is therefore perhaps surprising that we can glean some of the behaviour illustrated in the previous section---culminating in figure~\ref{figure-4}---from the case of an infinite number of layers.  Nevertheless, Bloch's theorem, which strictly only applies to a system of infinite size, can be used to accurately predict the region of high reflectivity in the central region of figure~\ref{figure-4} (10 layers).  An intuitive justification for its applicability is that for frequencies where the reflectivity is high, the wave is rapidly extinguished from the medium so that the field amplitude is very close to zero at the exit interface, and we can consider the number of unit cells to be infinite with little error.
%
%
	\par
	In an optical system with translational symmetry along an axis, \(x\), the eigenstates, \(\boldsymbol{E}\) transform as \(\boldsymbol{E}(x+\delta)=\boldsymbol{E}(x)e^{ik\delta}\), due to the fact that the translation operator commutes with the operators within the electromagnetic wave equation.  Bloch's theorem is the analogue of this relation for the case when we do not have continuous translational symmetry, but instead an infinite periodic structure, with discrete translational symmetry.  In this case, if the spatial periodicity is \(a+b\), then the eigenmodes obey,
	\begin{equation}
		\boldsymbol{E}(x+a+b)=\boldsymbol{E}(x)e^{i\kappa(a+b)}
	\end{equation}
	The quantity, \(\kappa\) is known as the Bloch wave--vector.  The Bloch wave--vector is the analogue of the free space wave--vector for the case of a discrete translational symmetry, and tells us the large scale variation of the field as one moves across one or more unit cells.  In terms of the transfer matrix formalism outlined in the previous two sections, this statement is,
	\begin{equation}
		\boldsymbol{\cal{M}}_{1}\boldsymbol{\cdot}\boldsymbol{e}=e^{i\kappa (a+b)}\boldsymbol{e},\label{bloch-theorem}	
	\end{equation}
	Finding the relation between the frequency, \(\omega\), and the Bloch-vector, \(\kappa\) amounts to finding the eigenvalues, \(\lambda=e^{i\kappa (a+b)}\) of the unit--cell matrix, \(\boldsymbol{\cal M}_{1}\).  These eigenvalues can be found in the usual manner, taking the right hand side of (\ref{bloch-theorem}) to the left and setting the determinant of the resulting matrix to zero,
	\[
		\text{det}\left(\boldsymbol{\cal{M}}_{1}-\lambda\boldsymbol{\mathds{1}}_{2}\right)=\left|\begin{matrix}{\cal{M}}_{11}-\lambda&{\cal{M}}_{12}\\{\cal{M}}_{21}&{\cal{M}}_{22}-\lambda\end{matrix}\right|=0,\label{eigenvalues}	
	\] 
	The quadratic equation obtained from (\ref{eigenvalues}) has the two solutions,
	\begin{equation}
		\lambda=z\pm\sqrt{z^{2}-1}=e^{i\kappa (a+b)}\label{eigenvalues-2}
	\end{equation}
	where we have applied \(\text{det}[\boldsymbol{\cal{M}}_{1}]=1\), and \(z=\text{Tr}[\boldsymbol{\cal{M}}_1]/2\).  These solutions determine the dispersion of the Bloch modes, \emph{i.e.}, the function \(\kappa(\omega)\).  It can be seen directly from (\ref{equation_9}) that \(z\) is real for media with \textit{real} refractive indices \(n_a\) and \(n_b\).  In this case the criterion for allowed propagation through the multilayer becomes quite simple, for when \(z\leq1\) we have,
	\[
		e^{i\kappa(a+b)}=z\pm i\sqrt{1-z^{2}}
	\]
	which can be satisfied with a real Bloch wave--vector: \(\kappa=\pm\arccos(z)/(a+b)\).  Meanwhile, when \(z>1\), the right hand side of (\ref{eigenvalues-2}) is real, with one root greater, and one less than unity.  This leads to a complex value of \(\kappa\).  Such solutions have divergent behaviour at infinity, and are therefore not allowed modes of the truly infinite system.  In a bounded medium these waves undergo \emph{extinction} inside the medium in the direction normal to the boundary. The decay is due to strong Bragg reflections, a reversible process that does not lead to a loss of energy from the field, making reflection within the gap strictly unity.
	\par
	In a \textit{dissipative} periodic medium (the dissipation being characterized by the difference between the incident and scattered power, \(1-R-T\)) the situation becomes rather intricate: \(z\) is complex.   To see how this modifies the situation, consider the case where a small degree of dissipation is present in an otherwise allowed frequency band: \(z=\cos(\varphi)+i\eta\), where both \(\varphi\) and \(\eta\) are real, and where \(\eta\ll1\).  To first order in \(\eta\) (\ref{eigenvalues-2}) becomes,
	\begin{equation}
		e^{i\kappa(a+b)}\sim \left(1\pm\frac{\eta}{\sin{(\varphi)}}\right)e^{\pm i\varphi}\label{solution-with-dissipation}
	\end{equation}
	The presence of dissipation evidently shifts the modulus of (\ref{solution-with-dissipation}) away from unity.  Therefore \(\kappa\) is in general complex when dissipation is present, even for modes falling within an otherwise allowed band.  Consequently the distinction between allowed and forbidden bands becomes blurred (see section~\ref{cold-trapped-atoms}).  This has the simple interpretation that in an absorbing periodic medium, the propagation of Bloch waves is damped.  This is a dissipative process, which prevents reflection from being unity.  For large degrees of dissipation the band-gap may even disappear.
%
%
\section{Some example applications\label{example-section}}
	\par
	So far we have given a treatment of the Bragg reflector in terms of transfer matrices, which might not be the standard presentation for undergraduates.  Otherwise the formulae are mostly well--known.  We shall now demonstrate how these results may be used to discuss aspects of physics that are relevant to some contemporary experimental set ups.
%
%
\subsection{A Bragg reflector of cold atoms loaded into an optical lattice\label{cold-trapped-atoms}}
%
%
\begin{figure}[h!]
\begin{center}
	\includegraphics[width=9cm]{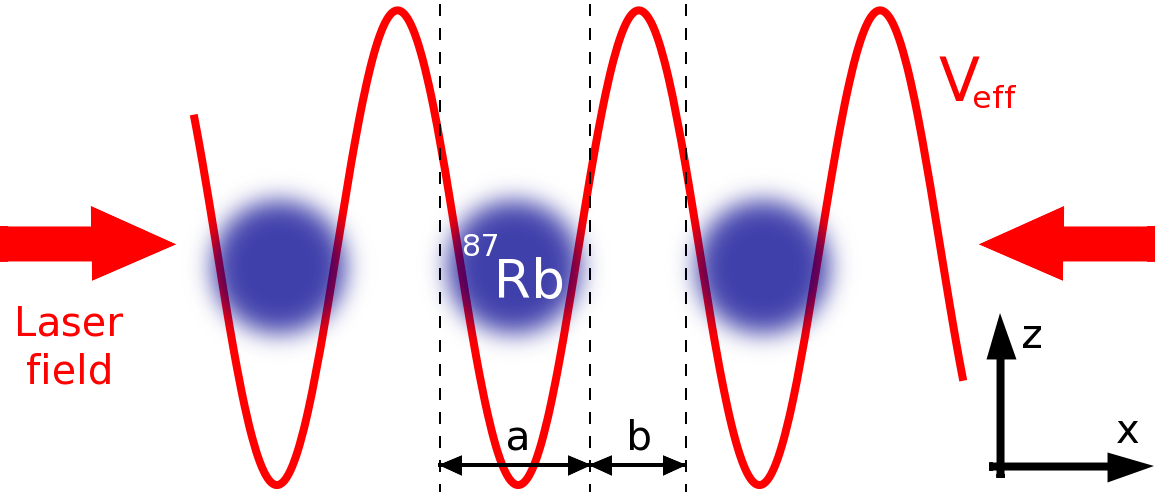}
		\caption{Neutral atoms can be trapped within the electric field of two counter--propagating red--detuned laser beams.  The electric field induces an atomic dipole moment, which interacts with the field, pushing the atomic centre of mass towards high field intensity.  When sufficiently cooled these atoms can be trapped so that they form a lattice.  A weak probe beam interacts with a 1D optical lattice in much the same way as a Bragg reflector composed of layers of vacuum and atomic vapour. \label{figure-5}}
\end{center}
\end{figure}
	\par
	A neutral atom subject to an external electric field will develop an electric dipole moment, \(\boldsymbol{d}\).  The atom will then be subject to a force, \(\boldsymbol{F}=d\boldsymbol{p}/dt\), due to interaction of the electromagnetic field with this induced polarization, and this force can be used to trap a cloud of such atoms within a confined region of space~\cite{grimm2000}.  In the situation illustrated in figure~\ref{figure-5}, the interference pattern created by two counter--propagating laser fields creates an array of many thousand of such traps, often termed an \emph{optical lattice}~\cite{bloch2005}.
	\par
	Consider a monochromatic field in which the induced atomic dipole moment can be written to first order in the electric field as, \(\boldsymbol{d}=\alpha(\omega)\boldsymbol{E}(\boldsymbol{r})\), where \(\alpha(\omega)\) is the single--atom polarizability (assumed real and positive, as appropriate for a laser field that is red--detuned from the atomic resonance), \(\boldsymbol{E}\) is the electric field amplitude (a three dimensional vector), and \(\boldsymbol{r}\) is the position of the atomic centre of mass.  The force on the centre of mass of each atom is, to first order in the size of the system,
	\[
		\frac{d\boldsymbol{p}}{dt}=\int\left[\rho\boldsymbol{E}+\boldsymbol{j}\boldsymbol{\times}\boldsymbol{B}\right]dV\sim\int\boldsymbol{x}\rho(\boldsymbol{x}+\boldsymbol{r})dV\boldsymbol{\cdot}\boldsymbol{\nabla}_{\boldsymbol{r}}\boldsymbol{E}(\boldsymbol{r})+\int\boldsymbol{j}(\boldsymbol{x}-\boldsymbol{r})dV\boldsymbol{\times}\boldsymbol{B}(\boldsymbol{r})
	\]
	where the integral is over the volume of the system of charges in the atom.  We assume that the velocity of the centre of mass is slow enough that we can neglect contributions to the force of order \(\dot{\boldsymbol{r}}/c\), and we also neglect the magnetic structure of the system.  Identifying, \(\boldsymbol{d}=\int\boldsymbol{x}\rho(\boldsymbol{x}+\boldsymbol{r})dV\), and \(\boldsymbol{\dot{d}}=\int\boldsymbol{j}(\boldsymbol{x}-\boldsymbol{r})dV\)\footnote{This can be shown from the continuity equation.  Multiply both sides by \(\boldsymbol{x}-\boldsymbol{r}\), and integrate over the system, \(\int(\boldsymbol{x}-\boldsymbol{r})\boldsymbol{\nabla}_{\boldsymbol{x}}\boldsymbol{\cdot}\boldsymbol{j}dV=-\int(\boldsymbol{x}-\boldsymbol{r})\frac{\partial\rho}{\partial t}dV\).  Integrating the left hand side by parts, we have the result, \(\int\boldsymbol{j}dV=\int(\boldsymbol{x}-\boldsymbol{r})\frac{\partial\rho}{\partial t}dV=\boldsymbol{\dot{d}}\).}, the force on the centre of mass can be written as, \(d\boldsymbol{p}/dt=(\boldsymbol{d}\boldsymbol{\cdot}\boldsymbol{\nabla}_{\boldsymbol{r}})\boldsymbol{E}(\boldsymbol{r})+\boldsymbol{\dot{d}}\boldsymbol{\times}\boldsymbol{B}\).  For the particular case of an induced dipole moment this then becomes,
	\begin{equation}
		\frac{d\boldsymbol{p}}{dt}=\frac{1}{2}\alpha(\omega)\boldsymbol{\nabla}|\boldsymbol{E}(\boldsymbol{r})|^{2}+\alpha(\omega)\frac{\partial}{\partial t}\left(\boldsymbol{E}\boldsymbol{\times}\boldsymbol{B}\right)\label{atom-force}
	\end{equation}
	Averaging the motion over a timescale significantly longer than the period of the optical field, the final term on the right of (\ref{atom-force}) becomes negligible.  The  force is such that the atomic centre of mass will be pulled towards regions of high electric field intensity.  In figure~\ref{figure-5} the red line indicates the effective potential, \(V_{\text{eff}}=-\frac{1}{2}\alpha(\omega)|\boldsymbol{E}(\boldsymbol{r})|^{2}\), in the minima of which we can imagine atoms collecting once their centre of mass motion has been cooled significantly by some suitable means~\cite{bardou2001}.
%
%
\begin{figure}[h!]
\begin{center}
	\subfloat[\label{figure-6a}]{\includegraphics[width=6.5cm]{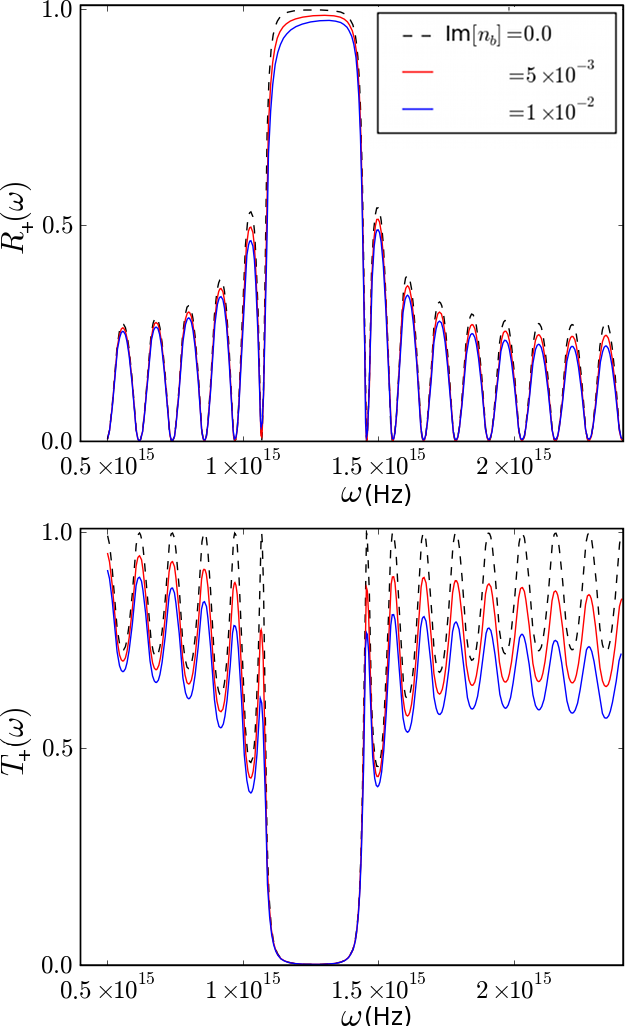}}\hspace{10pt}
	\subfloat[\label{figure-6b}]{\includegraphics[width=6.5cm]{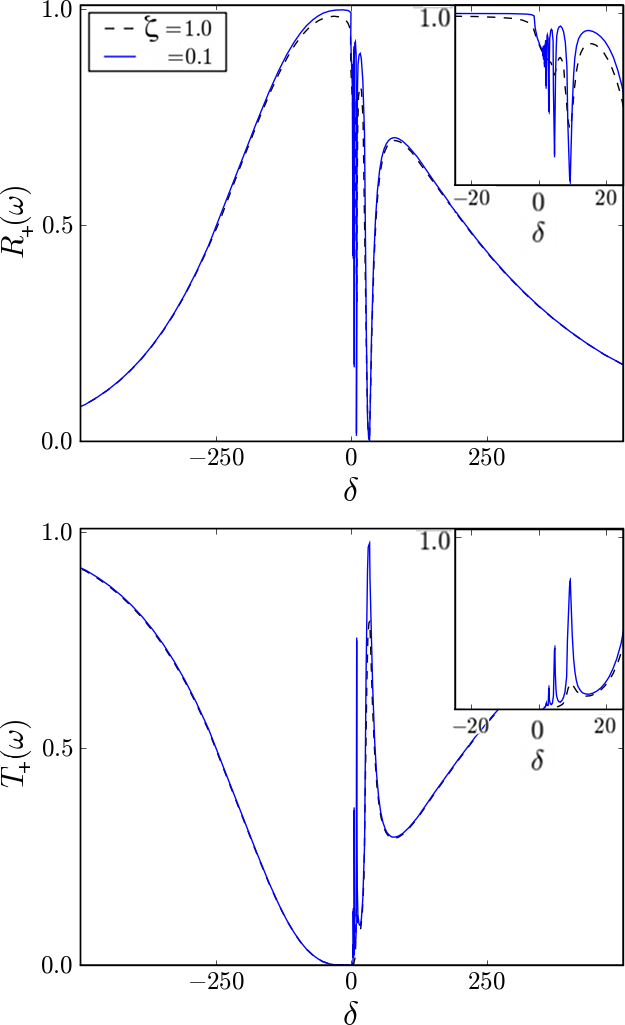}}
		\caption{Transmissivity \(T_{+}(\omega)\) and reflectivity \(R_{+}(\omega)\) as a function of frequency for a \textit{dissipative} Bragg reflector. (a) The dissipative analogue of the plot shown in figure~\ref{figure-4}.  All parameters are identical, but with the addition of an imaginary part to the refractive index, \(n_{b}\).  As \(\text{Im}[n_{b}]\) is increased, the contrast of the fringes around the stop band	 is decreased, and the shape of the reflectivity profile in the stop band changes. (b)  The array of unit cells now describes the index profile of cold atoms trapped within an optical lattice (\(\mathcal{N}=5.7\times10^{-3}\)).  The probe beam is assumed to be close to the frequency of the \(S_{1/2}\to P_{3/2}\) transition of \(^{87}\text{Rb}\), where the index can be approximated by (\ref{atomic-refractive-index}), with \(\omega_{0}=2.412\times10^{15}\,\text{Hz}\), \(\gamma_{e}=3.7699\times10^{7}\,\text{Hz}\), and we use the notation, \(\delta=(\omega_{0}-\omega)/\gamma_{e}\).  The trap has \(N=5.4\times10^{4}\) periods, and the periodicity is such that \(a=1.94807\times10^{-8}\,\text{m}\), and \(b=3.70873\times10^{-7}\,\text{m}\).  The blue and dashed lines illustrate the effect of changing the line shape parameter, \(\zeta\).\label{figure_6}}
\end{center}
\end{figure}
	\par	
	The optical lattice is useful as a model system for phenomena in condensed matter physics, but it also has interesting optical properties of its own~\cite{artoni2005,horsley2011}.  We can treat this system using the formalism developed in the previous sections (see figure~\ref{figure-5}).  In the case of a one dimensional optical lattice, where the atomic motion is only weakly constrained in the \(y-z\) plane, the system is essentially a slightly unusual Bragg reflector (an `atomic Bragg mirror').  The reflector is approximately composed of layers of atomic clouds (width \(a\), index \(n_{a}\)), and layers of vacuum (width \(b\), and \(n_{b,c}\sim1\)).
	\par
\begin{figure}[h!]
\begin{center}
	\subfloat[\label{figure-7a}]{\includegraphics[height=6.5cm]{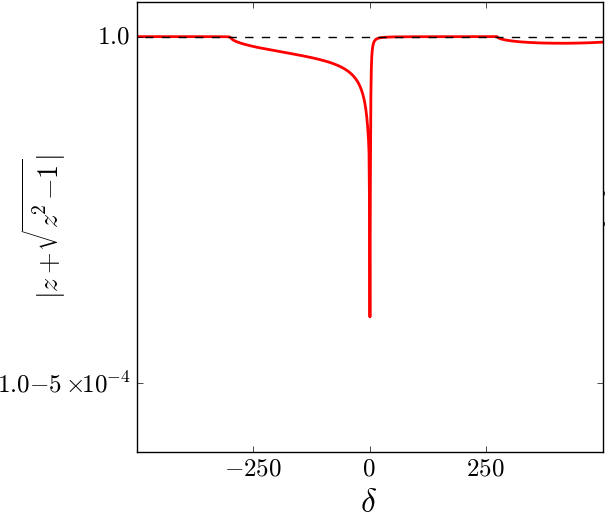}}\hspace{10pt}
	\subfloat[\label{figure-7b}]{\includegraphics[height=6.5cm]{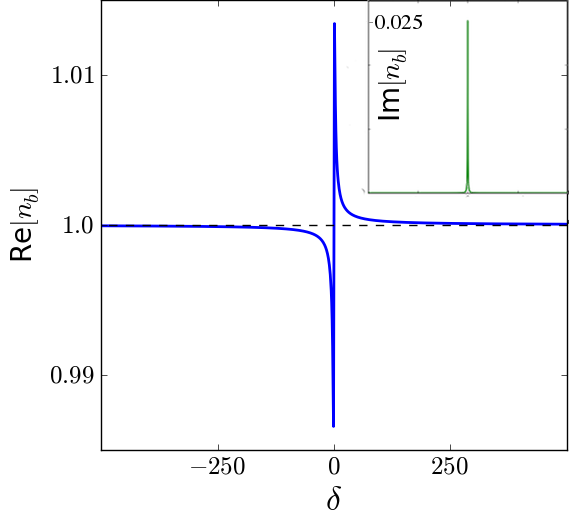}}
		\caption{(a) The absolute value of the right hand side of (\ref{eigenvalues-2}).  The deviation of this quantity away from unity indicates the magnitude of the imaginary part of the Bloch wave--vector. (b) The real and imaginary (inset) parts of the refractive index computed from (\ref{atomic-refractive-index}).  In both cases the parameters and variables are as in figure~\ref{figure-6b}\label{figure_7}}
\end{center}
\end{figure}

	Suppose that a weak beam of light probes the atoms; what are the reflection and transmission coefficients of the lattice?  For a weak probe each atom responds linearly to the electric field, and in the process develops an oscillating electric dipole moment, \(\boldsymbol{d}\).  Approximating this dipole moment as a simple harmonic motion of natural frequency \(\omega_{0}\), leads to \(\boldsymbol{d}\propto\boldsymbol{E}(\boldsymbol{x}_{0},t)/(\omega_{0}^{2}-\omega^{2}-i\omega\gamma)\), where \(\omega\) is the frequency of the applied electric field, \(\boldsymbol{x}_{0}\) is the position of the atom of interest, and \(\gamma\) represents the damping of the oscillatory motion.  For a cloud of such atoms, the macroscopic polarizability, \(\boldsymbol{P}=\chi(\omega)\boldsymbol{E}\) (average dipole moment per unit volume), has the same frequency dependence in response to a fixed amplitude electric field: \(\chi(\omega)\propto1/(\omega_{0}^{2}-\omega^{2}-i\omega\gamma)\).  The quantity, \(\chi(\omega)\) characterising the macroscopic polarization appears in the electric permittivity of the atomic vapour as, \(\epsilon=\epsilon_{0}+\chi(\omega)\), and can be related to the refractive index of the cloud through the formula, \(n=\sqrt{\epsilon/\epsilon_{0}}\).  Close to an atomic resonance, the denominator of \(\chi\) can be expanded to first order in \(\omega-\omega_{0}\) with little error, and the index of the atomic vapour thus takes the form,
	\begin{equation}
		n_{a}\left(\omega\right)\approx\sqrt{1+\frac{3\pi\mathcal{N}}{(\omega_{0}-\omega)/\gamma_{e}-i\zeta}}.\label{atomic-refractive-index}
	\end{equation}
	In (\ref{atomic-refractive-index}) the proportionality constant and the damping have been given specific values which come from a more rigorous quantum mechanical analysis~\cite{weiner2003}.  The quantity \(\gamma_{e}\) represents the linewidth of the excited state of the trapped atoms (i.e. the damping), which in the ideal case represents how broad the atomic resonance is in frequency.  The parameter \(\zeta\) is chosen to fit this width to the experimentally measured value, without changing the strength of the resonance (\(\zeta=1\) is the ideal theoretical value).  The strength of the resonance is proportional to \(\mathcal{N}\), which is the scaled atomic density, \(A/(k_{0}^{3}V)\) (\(A/V\)  atoms per unit volume), and \(k_{0}=\omega_{0}/c\).  The scaled atomic density measures the density of atoms relative to the wavelength of radiation at the resonant frequency.  The index (\ref{atomic-refractive-index}) is plotted in figure~\ref{figure_7}b.
%
%
	\par
	Figure~\ref{figure-6b} shows the reflection and transmission coefficients calculated from (\ref{transmission-result}--\ref{reflection-result}) and (\ref{atomic-refractive-index}), for the case when the trapped atoms are \(^{87}\text{Rb}\).  For the purposes of plotting we define the dimensionless parameter \(\delta=(\omega_{0}-\omega)/\gamma_{e}\), which measures how far the frequency of incident light is from resonance in units of the natural line width, and when positive corresponds to a red--detuning.  From the figure we can see that when \(\delta=0\), where the dissipation and dispersion are both most pronounced, the optical response is quite complicated, with rapid oscillations evident on the scale of \(\gamma_{e}\).  Either side of \(\delta=0\) \emph{two} bands of high reflectivity appear.  Figure~\ref{figure-7a} is a plot of the absolute value of the right hand of (\ref{eigenvalues-2}), illustrating the origin of these two bands.  For negative detuning away from resonance---where figure~\ref{figure-6b} shows nearly total reflection---the Bloch wave--vector has a significant imaginary part even though the absorption is small (inset of figure~\ref{figure-7b}).  There is also a stop band evident for positive detuning at around \(\delta\sim250\).  This is not completely evident in figure~\ref{figure-6b}, where the reflectivity peaks at \(\sim0.6\) for \(\delta\sim150\), but does become conspicuous for larger values of \(N\).
	\par
	This example system illustrates that when composed of resonant elements, the properties of the Bragg reflector can be quite subtle, and not at all like the textbook examples.  Furthermore it shows that even when the layers have an index contrast of less than a percent away from unity, it is still possible to create an almost perfect reflector.  This is so long as one can construct a system with \(N\) large enough, which is quite possible with an optical lattice~\cite{schilke2011,schilke2012}.
%
%
\subsection{Optomechanical cooling\label{optical-cooling}}
%
%
\begin{figure}[h!]
\begin{center}
	\includegraphics[width=8cm]{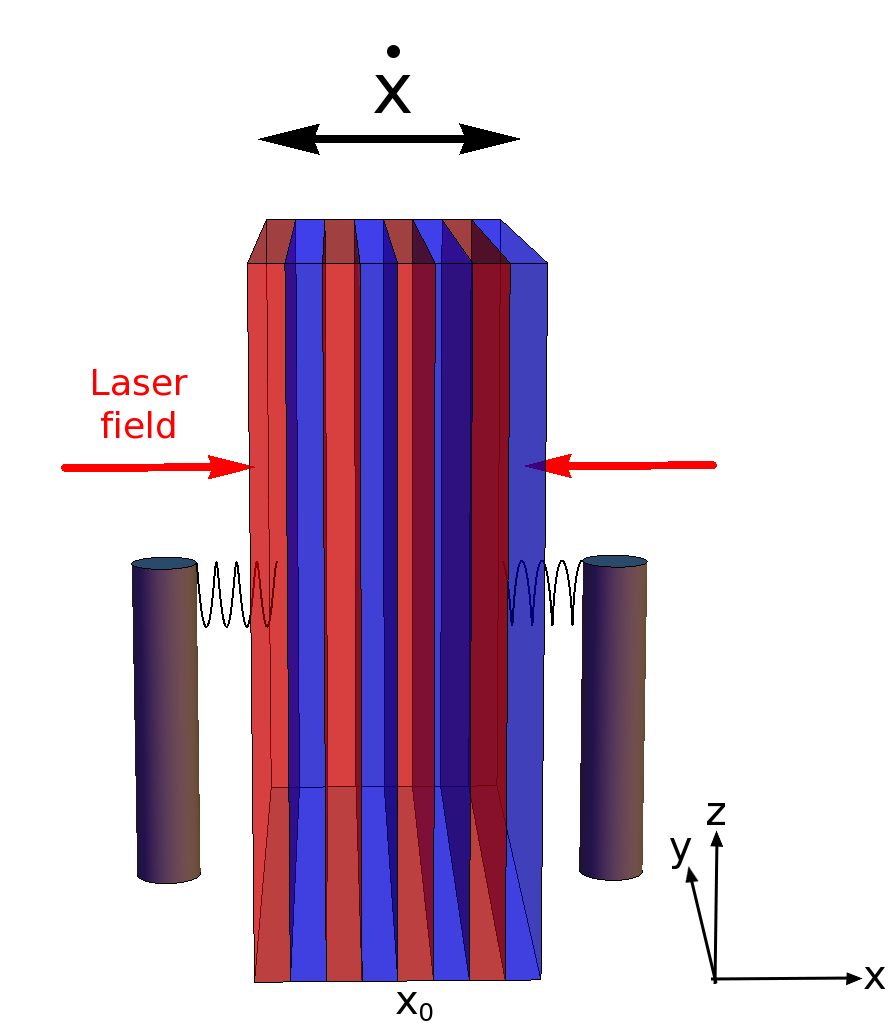}
	\caption{A Bragg reflector is allowed to move along the \(x\) axis, confined by a pair of springs attached to fixed columns, and performing an oscillatory motion around the point \(x=x_{0}\).  When equal amplitude, monochromatic light is incident onto both sides of the reflector, radiation pressure can be used to add or remove kinetic energy from the motion.\label{figure-8}}
\end{center}
\end{figure}  
	\par
	The electromagnetic field can add or remove energy from a mechanical degree of freedom.  In much the same way that atomic motion can be cooled within an optical lattice, macroscopic mechanical motion can be reduced to the point where it enters regime where a quantum mechanical description must be used~\cite{marquardt2009,aspelmeyer2010}.  For example, in the case of a mechanical oscillator of frequency, \(\Omega\), this would be when the total energy in the centre of mass motion becomes comparable with the spacing of the energy levels, \(\hbar\Omega\).  Recent experimental work has succeeded in reducing the vibrational energy of such a macroscopic degree of freedom to the quantum mechanical ground state~\cite{chan2011} using a process known as optomechanical cooling.  It is worth emphasising that this cooling refers to the macroscopic oscillatory motion and not the bulk temperature of the medium.    We consider the system illustrated in figure~\ref{figure-8} to demonstrate the basic principles (if not the detailed experimental set--up) involved in optomechanical cooling.  This is inspired by the fact that a similar set--up has been recently proposed for this purpose~\cite{karrai2008,horsley2011}.
	\par
	If monochromatic radiation is incident onto both sides of a Bragg reflector, due to the reflection from, and absorption into the body, it will exert a force.  For a single photon incident from the left (\(+\)) or right (\(-\)) with momentum \(\pm\hbar k_{\pm}\hat{\boldsymbol{x}}\) and energy \(\hbar\omega_{\pm}\), the difference in \emph{optical} energy (\(E^{F}\)) and momentum (\(\boldsymbol{p}^{F}\)) before and after interaction with the reflector would be on average,
	\begin{align}
		\Delta E^{F}_{\pm}&=\hbar \omega_{\pm}\left[T_{\pm}(\omega_{\pm})+R_{\pm}(\omega_{\pm})-1\right]\nonumber\\
		\Delta\boldsymbol{p}^{F}_{\pm}&=\pm\hbar k_{\pm}\left[T_{\pm}(\omega_{\pm})-R_{\pm}(\omega_{\pm})-1\right]\hat{\boldsymbol{x}}\label{left-lost}
	\end{align}	  
	where in terms of a single photon, \(T_{\pm}\) represents the probability that the photon passes straight through the medium, and \(R_{\pm}\) the probability that it is reflected, the quantity \(1-R_{\pm}-T_{\pm}\) is therefore the probability that it is lost from the field and absorbed by the body.  Assuming \(\mathcal{P}_{\pm}\) photons per second are incident from the left and right respectively, we can equate this rate of energy and momentum being lost from the field to that taken up by the body (i.e. the negative of (\ref{left-lost})).  The rest frame force on the reflector is then (a prime denoting the rest frame),
	\begin{equation}
		\frac{d\boldsymbol{p}^{\prime}_{\text{\tiny{EM}}}}{dt}=\hbar\left\{\mathcal{P}_{+} k_{+}\left[1+R(\omega_{+})-T(\omega_{+})\right]-\mathcal{P}_{-} k_{-}\left[1+R(\omega_{-})-T(\omega_{-})\right]\right\}\hat{\boldsymbol{x}}\label{rest-mom}
	\end{equation}
	and the energy transfer (bulk heating) is,
	\begin{equation}
		\frac{d E^{\prime}_{\text{\tiny{EM}}}}{dt}=\hbar\left\{\mathcal{P}_{+} \omega_{+}\left[1-R(\omega_{+})-T(\omega_{+})\right]+\mathcal{P}_{-} \omega_{-}\left[1-R(\omega_{-})-T(\omega_{-})\right]\right\}\label{rest-en}
	\end{equation}
	where we assume as is usually the case, that \(R_{+}\sim R_{-}=R\) and \(T_{+}=T_{-}=T\).
	\par
	In the laboratory, where the reflector is generally in motion with velocity \(\dot{x}\), we assume equal amplitude monochromatic radiation of frequency \(\omega\) incident from both directions.  To first order in the velocity, the rest frame frequency and wave--vector are given in terms of laboratory quantities by, \(\omega_{\pm}=\omega(1\mp\dot{x}/c)=c k_{\pm}\).  Furthermore, in the laboratory the number of photons sent towards the medium per second, \(\mathcal{P}_{0}\) is equal on both sides of the reflector, and this also transforms as a frequency, \(\mathcal{P}_{\pm}=\mathcal{P}_{0}(1\mp\dot{x}/c)\).  The rest frame exchange of momentum and energy (\ref{rest-mom}--\ref{rest-en}) are then,
	\begin{equation}
		\frac{d\boldsymbol{p}^{\prime}_{\text{\tiny{EM}}}}{dt}=-\frac{2\hbar\omega\mathcal{P}_{0}}{c}\left(\frac{\dot{x}}{c}\right)\left\{\omega\left[\frac{\partial R(\omega)}{\partial\omega}-\frac{\partial T(\omega)}{\partial\omega}\right]+2\left[1+R(\omega)-T(\omega)\right]\right\}\hat{\boldsymbol{x}}\label{final-rest-p}
	\end{equation}
	and,
	\begin{equation}
		\frac{d E^{\prime}_{\text{\tiny{EM}}}}{dt}=2\hbar\omega\mathcal{P}_{0}\left[1-R(\omega)-T(\omega)\right]\label{final-rest-E}
	\end{equation}
	where the reflection and transmission coefficients have been expanded as a Taylor series around the frequency of light as generated in the laboratory, \(R(\omega_{\pm})\sim R(\omega)\mp(\dot{x}\omega/c)\partial R(\omega)/\partial\omega\).  However, what we are interested in is the mechanical force on the body due to the electromagnetic field in the \emph{laboratory} frame of reference.  Transforming (\ref{final-rest-p}--\ref{final-rest-E}) from the rest frame into the laboratory frame, using \(d\boldsymbol{p}_{\text{\tiny{EM}}}/d t\sim d\boldsymbol{p}^{\prime}_{\text{\tiny{EM}}}/d t+(\dot{\boldsymbol{x}}/c^{2})d E^{\prime}_{\text{\tiny{EM}}}/d t\), the optical force on the reflector due to the field is,
	\begin{equation}
		\frac{d\boldsymbol{p}_{\text{\tiny{EM}}}}{dt}=-\frac{2\hbar\omega\mathcal{P}_{0}}{c}\left(\frac{\dot{x}}{c}\right)\left\{1+3R(\omega)-T(\omega)+\omega\left[\frac{\partial R(\omega)}{\partial\omega}-\frac{\partial T(\omega)}{\partial\omega}\right]\right\}\hat{\boldsymbol{x}}\label{optical-force}
	\end{equation}
	Expression (\ref{optical-force}) is the optical contribution to the force on a reflector, as observed moving at velocity \(\dot{x}\) in the laboratory frame.  Defining the following quantity,
	\begin{equation}
		\Gamma=\frac{\hbar\omega\mathcal{P}_{0}}{c^{2}}\left\{1+3R(\omega)-T(\omega)+\omega\left[\frac{\partial R(\omega)}{\partial\omega}-\frac{\partial T(\omega)}{\partial\omega}\right]\right\}\label{damping-coefficient}
	\end{equation}
	the equation of motion for the centre of mass of the Bragg reflector, attached to a pair of springs (figure~\ref{figure-8}), in the presence of the optical field is,
	\begin{equation}
		m\frac{d^{2}x}{dt^{2}}+2\Gamma\frac{d x}{dt}+m\Omega^{2}(x-x_{0})=0\label{damped-oscillator}
	\end{equation}
	where \(m\) is the mass of the reflector, and \(\Omega\) is the frequency associated with the restoring force of the springs.  The equation of motion, (\ref{damped-oscillator}) is of the form of a damped (\(\Gamma>0\)), or amplified (\(\Gamma<0\)) harmonic oscillator.  The damping coefficient is proportional to the incident optical power, and depends on the reflection and transmission coefficients and their derivatives with respect to frequency.  The solutions to (\ref{damped-oscillator}) have the form,
	\[
		x(t)=x_{0}+A_{0} e^{-\Gamma t/m}e^{\pm i\Omega\sqrt{1-\Gamma^{2}/m^{2}\Omega^{2}}t}
	\]
	where \(A_{0}\) is the distance of the centre of mass from \(x_{0}\) at \(t=0\).  If the medium is only weakly dispersive, then the derivatives of the reflection and transmission coefficients within (\ref{damping-coefficient}) may be neglected: then \(\Gamma\) is strictly positive.  Without such terms it is evident that the rate of damping is ordinarily very small.  For instance, an optical field, \(\omega\sim10^{15}\,\text{Hz}\), and a reflector of a very small mass~\cite{karrai2008}, \(m\sim10^{-15}\,\text{kg}\), yields, \(\Gamma/m\sim10^{-23}\mathcal{P}_{0}\).  For a moderate damping rate of \(\Gamma/m\sim10^{-3}\) we would thus require \(\mathcal{P}_{0}\sim10^{20}\), corresponding to a laser power per unit area of \(10\,\text{W}/\mathcal{A}\), where \(\mathcal{A}\) is the cross sectional area of the reflector.  In such a field the heating due to absorption alone would be enough to melt the medium.
%
%
\begin{figure}
	\includegraphics[width=10cm]{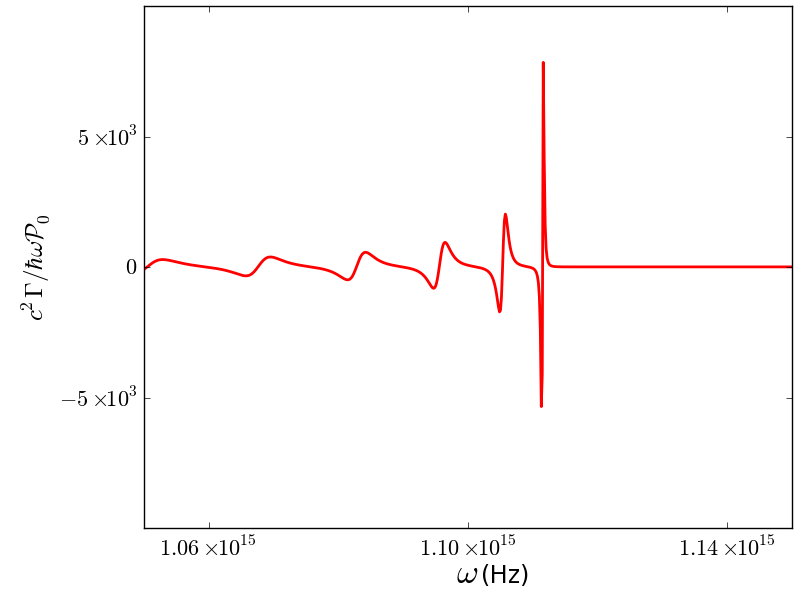}
	\caption{The damping coefficient, \(\Gamma\) in units of \(\hbar\omega\mathcal{P}_{0}/c^{2}\), for a Bragg reflector in a frequency regime around the edge of a stop band.  The parameters are as in figure~\ref{figure-4}, but for the case of \(N=50\) layers, where the stop band is more defined.\label{figure-9}}
\end{figure}
	\par
	Conversely for a Bragg reflector, dispersion is not negligible.  Figure~\ref{figure-4} shows there are strong oscillations in the reflection and transmission coefficients, contributing significantly to \(\Gamma\) via the frequency derivatives in (\ref{damping-coefficient}).  The motion can then be either amplified or damped (cooled), depending on the sign of the derivatives of the reflection and transmission coefficients.  For small dissipation, and increasing reflectivity with frequency we have damping (\(\Gamma>0\)), while for decreasing reflectivity with frequency, we have amplification (\(\Gamma<0\)).  Figure~\ref{figure-9} illustrates \(c^{2}\Gamma/\hbar\omega\mathcal{P}_{0}\), demonstrating four orders of magnitude increase in the absolute value of the damping coefficient.  The required optical power density is correspondingly four orders of magnitude smaller.  The above effect is the basic principle of optomechanical cooling.
	\par
	It is worth mentioning that while the Bragg reflector is highly dispersive, a typical experiment in optomechanics involves an object that has relatively weak dispersion.  However, the object sits within a high quality factor optical cavity.  The frequency of incident radiation is close to a cavity resonance, which is the source of the strong dispersion in the equivalent of \(\Gamma\).
%
%
\subsection{Optical non-reciprocity\label{non-reciprocity-sec}}	
	\par
	Reciprocity---the fact that the same field will be detected when the positions of a source and detector of waves are interchanged---is a rather general property of wave propagation.  For the purposes of performing signal processing, or computation with photons, one might like to realise an optical diode, allowing total light transmission in the forward direction, and inhibiting (over some bandwidth) propagation in the backward direction.  Partly for this reason, current efforts are going into the realisation of non--reciprocal optical devices.  A moving Bragg reflector provides a simple example of a system exhibiting significant non-reciprocity (due to the motion, which breaks time symmetry).  Indeed close to the stop band edge its properties can be much like that of an optical diode.
%
%
\begin{figure}[h!]
	\begin{center}
	\includegraphics[width=12cm]{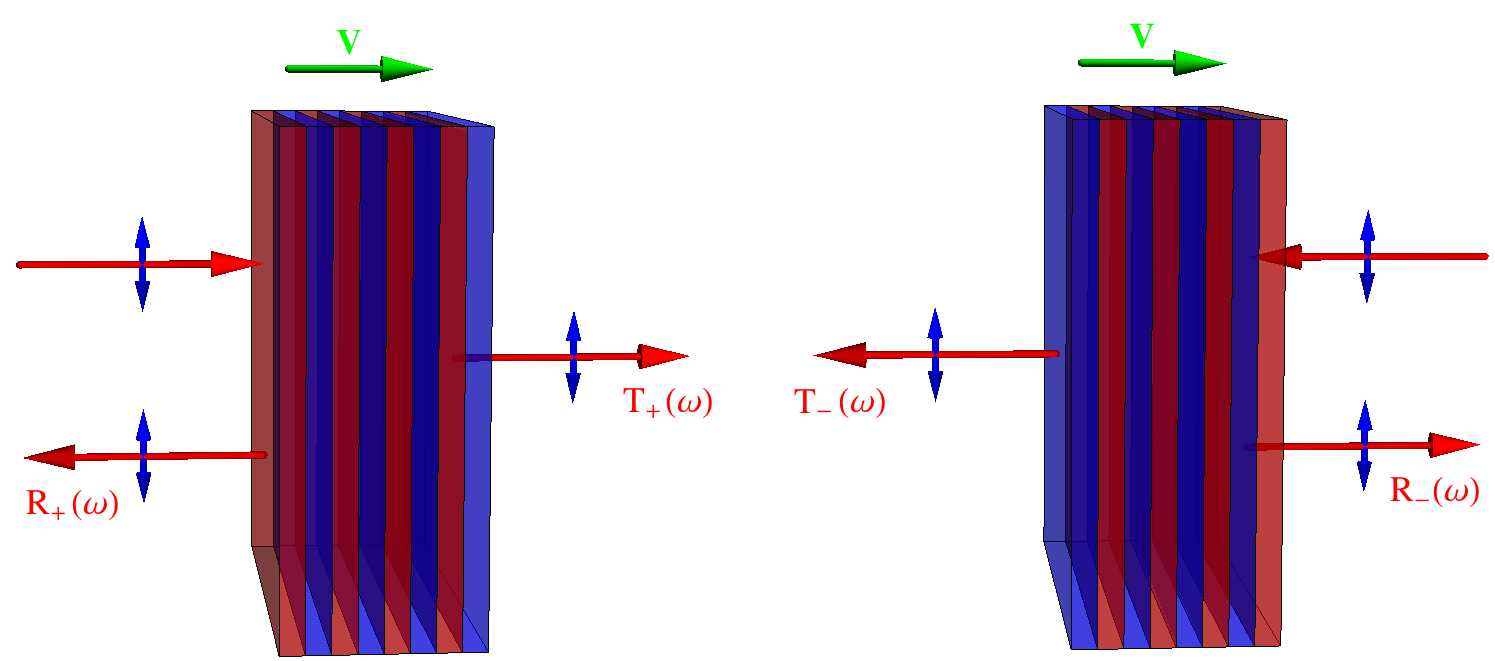}
	\caption{In the laboratory frame a Bragg reflector moves with velocity \(\boldsymbol{v}\) along \(\hat{\boldsymbol{x}}\).  Light of frequency \(\omega\), incident on the multilayer structure, is seen to exhibit significantly different reflection and transmission characteristics, depending on whether it impinges from the right \{$R_{-}, T_{-}$\}, or from the left \{$R_{+}, T_{+}$\}.\label{figure-10}}
	\end{center}
	\end{figure}
	\par
	Consider propagation through an arbitrary layered medium, as described in section~\ref{tm-section}, and illustrated in figure~\ref{figure_1}.  If the medium is reciprocal, then the transmission from the background index on the left of the medium to that on the right should be the same as transmission from right to left.  We shall define a simple measure of the non--reciprocity of our layered medium to be,
\begin{equation}
	\Delta T=T_{+}(\omega)-T_{-}(\omega).\label{non-reciprocity}
\end{equation}
	where the transmission coefficients are those measured in the laboratory.  This measure obviously doesn't capture everything that the full definition of non--reciprocity does~\cite{volume8}, but should be in qualitative agreement.  Computing \(T_{+}\) and \(T_{-}\) as in section~\ref{bragg-section}, one obtains \(T_{+}=|\text{det}(\boldsymbol{M})/m_{22}|^{2}\), and \(T_{-}=|1/m_{22}|^{2}\) so that for \(T_{+}\) to equal \(T_{-}\), it is evident that we must have \(\text{det}[\boldsymbol{M}]=1\).  For normal incidence onto a non--magnetic medium, an arbitrary transfer matrix is equal to a (possibly infinite) product of \(\boldsymbol{X}\) and \(\boldsymbol{W}\) matrices given by (\ref{equation_2}--\ref{equation_5}), and the determinant of such a product equals the product of the determinants,
	\begin{align}
		\text{det}[\boldsymbol{M}]&=\prod_{i=1}^{N}\text{det}[\boldsymbol{X}^{(i)}(d_{i})]\prod_{j=0}^{N}\text{det}[\boldsymbol{W}^{(j,j+1)}]\nonumber\\
		&=\prod_{j=0}^{N}\frac{n_{j}}{n_{j+1}}=\frac{n_{0}}{n_{N+1}}\label{det-trans}
	\end{align}
	where we have assumed an \(N\) layered medium, each layer having index \(n_{i}\) and thickness \(d_{i}\).  As defined, (\ref{non-reciprocity}) is applicable only to the case where \(n_{0}=n_{N+1}\), and thus \(T_{+}=T_{-}\).  In the case where source and detector reside in different media, the simple measure (\ref{non-reciprocity}) will not even be in qualitative agreement with that found in e.g. Landau and Lifshitz~\cite{volume8}, and must be modified.
%
%
\begin{figure}
	\includegraphics[width=10cm]{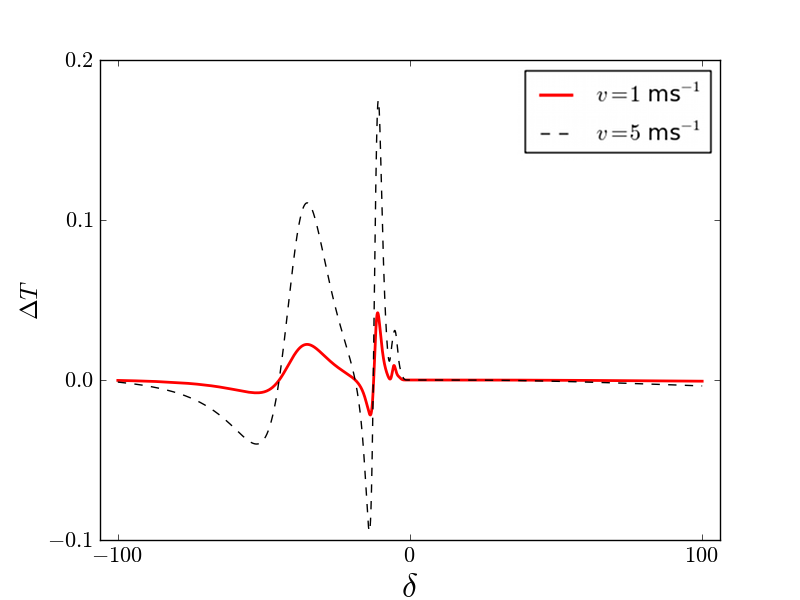}
	\caption{Non reciprocal response, (\ref{non-reciprocity}) for an atomic Bragg mirror set in motion.  The velocity is \(1\,\text{ms}^{-1}\) (solid red) and \(5\,\text{ms}^{-1}\) (black dashed).  The parameters are those given in figure~\ref{figure-6b}.\label{figure-11}}
\end{figure}
	\par
	Reciprocity is therefore necessarily embedded within the \(2\times2\) transfer matrix formalism.  To break reciprocity we must do something to break the formalism.  Perhaps the simplest way to do this is to take a medium that has been set into motion, as in figure~\ref{figure-10}.  We cannot then simply take \(\omega\) to be the same throughout the system.  Reflection causes the frequency of the waves to become \(\omega_{\pm}=\omega(1\mp v/c)\), and this will depend on the direction of incidence.  Therefore, incident left and right going waves of frequency \(\omega\) will generate reflected and transmitted waves of frequencies, \(\omega_{\pm}\) and \(\omega\): for two input channels we have four output channels.
	\par
	To calculate (\ref{non-reciprocity}) for our system, consider propagation in the rest frame.  For incidence from the left, the rest frame field on either side of the medium is
	\begin{equation}
		\boldsymbol{E}^{\prime}(x^{\prime})=\begin{cases}E_{+}\hat{\boldsymbol{z}}\left[e^{i \omega_{+}x^{\prime}/c}+r_{+}(\omega_{+})e^{-i \omega_{+}x^{\prime}/c}\right]&x^{\prime}<0\\
		E_{+}\hat{\boldsymbol{z}}t(\omega_{+}) e^{i \omega_{+}x^{\prime}/c}&x^{\prime}>L
		\end{cases}\label{left-inc}
	\end{equation}
	and for incidence from the right,
	\begin{equation}
		\boldsymbol{E}^{\prime}(x^{\prime})=\begin{cases}E_{-}\hat{\boldsymbol{z}}t (\omega_{-})e^{-i \omega_{-}x^{\prime}/c}&x^{\prime}<0\\
		E_{-}\hat{\boldsymbol{z}}\left[e^{-i \omega_{-}x^{\prime}/c}+r_{-}(\omega_{-})e^{i \omega_{-}x^{\prime}/c}\right]&x^{\prime}>L
		\end{cases}\label{right-inc}
	\end{equation}
	where the medium is taken to be of length \(L\).  Expressions (\ref{left-inc}--\ref{right-inc}) are transformed into the laboratory frame using the first order transformation, \(\boldsymbol{E}=\boldsymbol{E}^{\prime}-\boldsymbol{v}\boldsymbol{\times}\boldsymbol{B}^{\prime}\),
	\begin{equation}
		\boldsymbol{E}(x)=\begin{cases}
						E_{0}\hat{\boldsymbol{z}}\left[e^{i\omega x/c}+(1-2v/c)r_{+}(\omega_{+})e^{-i\omega x/c}\right]&x<vt\\
						E_{0}\hat{\boldsymbol{z}}t(\omega_{+})e^{i\omega x/c}&x>L+vt
		\end{cases}\label{lab-1}
	\end{equation}
	and,
	\begin{equation}
		\boldsymbol{E}(x)=\begin{cases}
						E_{0}\hat{\boldsymbol{z}}(1+v/c)t(\omega_{-})e^{-i\omega x/c}&x<vt\\
						E_{0}\hat{\boldsymbol{z}}(1+v/c)\left[e^{-i\omega x/c}+(1+2v/c)r_{-}(\omega_{-})e^{i\omega x/c}\right]&x>L+vt
		\end{cases}\label{lab-2}
	\end{equation}
	where \(E_{+}=E_{0}(1-v/c)\) and \(E_{-}=E_{0}(1+v/c)\), the notation being as in section~\ref{optical-cooling}.  The difference in laboratory frame transmissivities for left and right incidence can thus be inferred from (\ref{lab-1}--\ref{lab-2}) and simply involves the difference in the rest frame transmission coefficients for two different Doppler shifted frequencies,
	\begin{equation}
		\Delta T=T_{+}-T_{-}=|t(\omega(1-v/c))|^{2}-|t(\omega(1+v/c))|^{2}\label{norec}
	\end{equation}
	To first order in the velocity of the medium, (\ref{norec}) is,
	\begin{equation}
		\Delta T\sim-\frac{2\omega v}{c}\frac{\partial T(\omega)}{\partial\omega}\label{delta-T}
	\end{equation}
	where \(T(\omega)\) is the rest frame transmissivity.  If dispersion is moderate then this measure of non--reciprocity is negligibly small for realistic velocities (say \(T\) changes \(0.1\) over \(10^{15}\,\text{Hz}\) then for optical frequencies and \(v\sim1\,\text{ms}^{-1}\), \(\Delta T\sim10^{-9}\)).  However, when the rest frame transmission coefficient depends strongly on frequency, then (\ref{norec}) may be large: again the Bragg reflector illustrates a prime example of this.  Figure~\ref{figure-11} gives a plot of \(\Delta T\) for an atomic Bragg mirror (section~\ref{cold-trapped-atoms}) set into motion\footnote{An atomic Bragg mirror may be set into motion by relatively detuning the two laser beams in figure~\ref{figure-5} so that the interference pattern moves along at a velocity determined by the detuning.  Alternatively the lattice may be given a sudden jolt so that the atoms slosh about within the wells.  In this latter case the velocity of the medium is a periodic function of time.} at \(1\,\text{ms}^{-1}\) and \(5\,\text{ms}^{-1}\).  Due to the strongly dispersive optical response of the lattice (figure~\ref{figure-6b}) the non--reciprocity of this system as determined by \(\Delta T\) can be in the tens of percent for velocities of metres per second.  For the same velocity regime, an effective optical diode exhibiting nearly \(100\%\) isolation would require only an order of magnitude increase in the dispersion.
%
%
\section{Conclusions}
	\par
	We have given an overview of the properties of the Bragg reflector using a simple application of the transfer matrix formalism.  The basic physics of this system was found to be convenient for discussing several topical areas of optical physics.  We used it, in particular, as the basis for a discussion of optomechanical cooling and optical nonÐreciprocity effects which can all be observed by using Bragg mirrors made of ultra-cold atoms.
%
%
\acknowledgements
SARH thanks the EPSRC for financial support.  This work is supported by the National Natural Science Foundation of China (11104112), the National Basic Research Program of China (2011CB921603), the
CRUI-British Council Programs \textquotedblleft Atoms and Nanostructures" and \textquotedblleft Metamaterials", and the IT09L244H5 Azione Integrata MIUR grant and the 2011 Fondo di Ateneo of Brescia University. Two of us (MA and GLR) would like to thank J.-H. Wu for the hospitality at Jilin.
%
%
\bibliography{simon-ajp}

\begin{thebibliography}{26}
\expandafter\ifx\csname natexlab\endcsname\relax\def\natexlab#1{#1}\fi
\expandafter\ifx\csname bibnamefont\endcsname\relax
  \def\bibnamefont#1{#1}\fi
\expandafter\ifx\csname bibfnamefont\endcsname\relax
  \def\bibfnamefont#1{#1}\fi
\expandafter\ifx\csname citenamefont\endcsname\relax
  \def\citenamefont#1{#1}\fi
\expandafter\ifx\csname url\endcsname\relax
  \def\url#1{\texttt{#1}}\fi
\expandafter\ifx\csname urlprefix\endcsname\relax\def\urlprefix{URL }\fi
\providecommand{\bibinfo}[2]{#2}
\providecommand{\eprint}[2][]{\url{#2}}

\bibitem[{\citenamefont{Griffiths and Steinke}(2001)}]{griffiths2001}
\bibinfo{author}{\bibfnamefont{D.~J.} \bibnamefont{Griffiths}}
  \bibnamefont{and} \bibinfo{author}{\bibfnamefont{C.~A.}
  \bibnamefont{Steinke}}, \bibinfo{journal}{Am. J. Phys.}
  \textbf{\bibinfo{volume}{69}}, \bibinfo{pages}{137} (\bibinfo{year}{2001}).

\bibitem[{\citenamefont{Ziman}(1972)}]{ziman1972}
\bibinfo{author}{\bibfnamefont{J.~M.} \bibnamefont{Ziman}},
  \emph{\bibinfo{title}{Principles of the theory of solids}}
  (\bibinfo{publisher}{Cambridge University Press},
  \bibinfo{address}{Cambridge, UK}, \bibinfo{year}{1972}).

\bibitem[{\citenamefont{Lifshitz and Pitaevskii}(2003)}]{volume9}
\bibinfo{author}{\bibfnamefont{E.~M.} \bibnamefont{Lifshitz}} \bibnamefont{and}
  \bibinfo{author}{\bibfnamefont{L.~P.} \bibnamefont{Pitaevskii}},
  \emph{\bibinfo{title}{Statistical Physics (Part 2)}}
  (\bibinfo{publisher}{Butterworth-Heinemann}, \bibinfo{year}{2003}).

\bibitem[{\citenamefont{Olsen and Vignale}(2010)}]{olsen2010}
\bibinfo{author}{\bibfnamefont{R.~J.} \bibnamefont{Olsen}} \bibnamefont{and}
  \bibinfo{author}{\bibfnamefont{G.}~\bibnamefont{Vignale}},
  \bibinfo{journal}{Am. J. Phys.} \textbf{\bibinfo{volume}{78}},
  \bibinfo{pages}{954} (\bibinfo{year}{2010}).

\bibitem[{\citenamefont{Born and Wolf}(1999)}]{born1999}
\bibinfo{author}{\bibfnamefont{M.}~\bibnamefont{Born}} \bibnamefont{and}
  \bibinfo{author}{\bibfnamefont{E.}~\bibnamefont{Wolf}},
  \emph{\bibinfo{title}{Principles of Optics}} (\bibinfo{publisher}{Cambridge
  University Press}, \bibinfo{year}{1999}).

\bibitem[{\citenamefont{Griffiths and Taussig}(1992)}]{griffiths1992}
\bibinfo{author}{\bibfnamefont{D.}~\bibnamefont{Griffiths}} \bibnamefont{and}
  \bibinfo{author}{\bibfnamefont{N.}~\bibnamefont{Taussig}},
  \bibinfo{journal}{Am. J. Phys.} \textbf{\bibinfo{volume}{60}},
  \bibinfo{pages}{883} (\bibinfo{year}{1992}).

\bibitem[{\citenamefont{Guo}(2006)}]{guo2006}
\bibinfo{author}{\bibfnamefont{W.}~\bibnamefont{Guo}}, \bibinfo{journal}{Am. J.
  Phys.} \textbf{\bibinfo{volume}{74}}, \bibinfo{pages}{595}
  (\bibinfo{year}{2006}).

\bibitem[{\citenamefont{S\'anchez-Soto
  et~al.}(2012)\citenamefont{S\'anchez-Soto, Monz\`on, Barriuso, and
  Cari\~nena}}]{soto2012}
\bibinfo{author}{\bibfnamefont{L.~L.} \bibnamefont{S\'anchez-Soto}},
  \bibinfo{author}{\bibfnamefont{J.~J.} \bibnamefont{Monz\`on}},
  \bibinfo{author}{\bibfnamefont{A.~G.} \bibnamefont{Barriuso}},
  \bibnamefont{and} \bibinfo{author}{\bibfnamefont{J.~F.}
  \bibnamefont{Cari\~nena}}, \bibinfo{journal}{Phys. Rep.}
  \textbf{\bibinfo{volume}{513}}, \bibinfo{pages}{191} (\bibinfo{year}{2012}).

\bibitem[{\citenamefont{Bloch}(2005)}]{bloch2005}
\bibinfo{author}{\bibfnamefont{I.}~\bibnamefont{Bloch}}, \bibinfo{journal}{Nat.
  Phys.} \textbf{\bibinfo{volume}{1}}, \bibinfo{pages}{23}
  (\bibinfo{year}{2005}).

\bibitem[{\citenamefont{Marquardt and Girvin}(2009)}]{marquardt2009}
\bibinfo{author}{\bibfnamefont{F.}~\bibnamefont{Marquardt}} \bibnamefont{and}
  \bibinfo{author}{\bibfnamefont{S.~M.} \bibnamefont{Girvin}},
  \bibinfo{journal}{Physics} \textbf{\bibinfo{volume}{2}}, \bibinfo{pages}{40}
  (\bibinfo{year}{2009}).

\bibitem[{\citenamefont{Potton}(2004)}]{potton2004}
\bibinfo{author}{\bibfnamefont{R.~J.} \bibnamefont{Potton}},
  \bibinfo{journal}{Rep. Prog. Phys.} \textbf{\bibinfo{volume}{67}},
  \bibinfo{pages}{717} (\bibinfo{year}{2004}).

\bibitem[{\citenamefont{Dai et~al.}(2012)\citenamefont{Dai, Bauters, and
  Bowers}}]{dai2012}
\bibinfo{author}{\bibfnamefont{D.}~\bibnamefont{Dai}},
  \bibinfo{author}{\bibfnamefont{J.}~\bibnamefont{Bauters}}, \bibnamefont{and}
  \bibinfo{author}{\bibfnamefont{J.~E.} \bibnamefont{Bowers}},
  \bibinfo{journal}{Nat. Light Sci. \& App.} \textbf{\bibinfo{volume}{1}},
  \bibinfo{pages}{e1} (\bibinfo{year}{2012}).

\bibitem[{\citenamefont{Landau et~al.}(2004)\citenamefont{Landau, Lifshitz, and
  Pitaevskii}}]{volume8}
\bibinfo{author}{\bibfnamefont{L.~D.} \bibnamefont{Landau}},
  \bibinfo{author}{\bibfnamefont{E.~M.} \bibnamefont{Lifshitz}},
  \bibnamefont{and} \bibinfo{author}{\bibfnamefont{L.~P.}
  \bibnamefont{Pitaevskii}}, \emph{\bibinfo{title}{The Electrodynamics of
  Continuous Media}} (\bibinfo{publisher}{{Butterworth-Heinemann}},
  \bibinfo{address}{Oxford}, \bibinfo{year}{2004}).

\bibitem[{\citenamefont{Heavens}(1960)}]{heavens1960}
\bibinfo{author}{\bibfnamefont{O.~S.} \bibnamefont{Heavens}},
  \bibinfo{journal}{Rep. Prog. Phys.} \textbf{\bibinfo{volume}{23}},
  \bibinfo{pages}{1} (\bibinfo{year}{1960}).

\bibitem[{\citenamefont{Yeh}(2005)}]{yeh2005}
\bibinfo{author}{\bibfnamefont{P.}~\bibnamefont{Yeh}},
  \emph{\bibinfo{title}{Optical Waves in Layered Media}}
  (\bibinfo{publisher}{Wiley}, \bibinfo{year}{2005}).

\bibitem[{\citenamefont{Gradshteyn and Ryzhik}(2000)}]{gradshteyn2000}
\bibinfo{author}{\bibfnamefont{I.~S.} \bibnamefont{Gradshteyn}}
  \bibnamefont{and} \bibinfo{author}{\bibfnamefont{I.~M.}
  \bibnamefont{Ryzhik}}, \emph{\bibinfo{title}{Table of Integrals, Series and
  Products}} (\bibinfo{publisher}{Academic Press}, \bibinfo{address}{San Diego,
  USA}, \bibinfo{year}{2000}).

\bibitem[{\citenamefont{Grimm et~al.}(2000)\citenamefont{Grimm, Weidem\"uller,
  and Ovchinnikov}}]{grimm2000}
\bibinfo{author}{\bibfnamefont{R.}~\bibnamefont{Grimm}},
  \bibinfo{author}{\bibfnamefont{M.}~\bibnamefont{Weidem\"uller}},
  \bibnamefont{and} \bibinfo{author}{\bibfnamefont{Y.~B.}
  \bibnamefont{Ovchinnikov}}, \bibinfo{journal}{Adv. At. Mol. Opt. Phys.}
  \textbf{\bibinfo{volume}{42}}, \bibinfo{pages}{95} (\bibinfo{year}{2000}).

\bibitem[{\citenamefont{Bardou et~al.}(2001)\citenamefont{Bardou, Bouchaud,
  Aspect, and Cohen-Tannoudji}}]{bardou2001}
\bibinfo{author}{\bibfnamefont{F.}~\bibnamefont{Bardou}},
  \bibinfo{author}{\bibfnamefont{J.-P.} \bibnamefont{Bouchaud}},
  \bibinfo{author}{\bibfnamefont{A.}~\bibnamefont{Aspect}}, \bibnamefont{and}
  \bibinfo{author}{\bibfnamefont{C.}~\bibnamefont{Cohen-Tannoudji}},
  \emph{\bibinfo{title}{L\'evy Statistics and Laser Cooling: How Rare Events
  Bring Atoms to Rest}} (\bibinfo{publisher}{Cambridge University Press},
  \bibinfo{address}{Cambridge}, \bibinfo{year}{2001}).

\bibitem[{\citenamefont{Artoni et~al.}(2005)\citenamefont{Artoni, La~Rocca, and
  Bassani}}]{artoni2005}
\bibinfo{author}{\bibfnamefont{M.}~\bibnamefont{Artoni}},
  \bibinfo{author}{\bibfnamefont{G.~C.} \bibnamefont{La~Rocca}},
  \bibnamefont{and} \bibinfo{author}{\bibfnamefont{F.}~\bibnamefont{Bassani}},
  \bibinfo{journal}{Phys. Rev. E} \textbf{\bibinfo{volume}{72}},
  \bibinfo{pages}{046604} (\bibinfo{year}{2005}).

\bibitem[{\citenamefont{Horsley et~al.}(2011)\citenamefont{Horsley, Artoni, and
  La~Rocca}}]{horsley2011}
\bibinfo{author}{\bibfnamefont{S.~A.~R.} \bibnamefont{Horsley}},
  \bibinfo{author}{\bibfnamefont{M.}~\bibnamefont{Artoni}}, \bibnamefont{and}
  \bibinfo{author}{\bibfnamefont{G.~C.} \bibnamefont{La~Rocca}},
  \bibinfo{journal}{Phys. Rev. Lett.} \textbf{\bibinfo{volume}{107}},
  \bibinfo{pages}{043602} (\bibinfo{year}{2011}).

\bibitem[{\citenamefont{Weiner and Ho}(2003)}]{weiner2003}
\bibinfo{author}{\bibfnamefont{J.}~\bibnamefont{Weiner}} \bibnamefont{and}
  \bibinfo{author}{\bibfnamefont{P.-T.} \bibnamefont{Ho}},
  \emph{\bibinfo{title}{Light-Matter Interaction: Fundamentals and Applications
  (Volume 1)}} (\bibinfo{publisher}{Wiley-Interscience}, \bibinfo{address}{New
  Jersey}, \bibinfo{year}{2003}).

\bibitem[{\citenamefont{Schilke et~al.}(2011)\citenamefont{Schilke, Zimmermann,
  Courteille, and Guerin}}]{schilke2011}
\bibinfo{author}{\bibfnamefont{A.}~\bibnamefont{Schilke}},
  \bibinfo{author}{\bibfnamefont{C.}~\bibnamefont{Zimmermann}},
  \bibinfo{author}{\bibfnamefont{P.~W.} \bibnamefont{Courteille}},
  \bibnamefont{and} \bibinfo{author}{\bibfnamefont{W.}~\bibnamefont{Guerin}},
  \bibinfo{journal}{Phys. Rev. Lett.} \textbf{\bibinfo{volume}{106}},
  \bibinfo{pages}{223903} (\bibinfo{year}{2011}).

\bibitem[{\citenamefont{Schilke et~al.}(2012)\citenamefont{Schilke, Zimmermann,
  and Guerin}}]{schilke2012}
\bibinfo{author}{\bibfnamefont{A.}~\bibnamefont{Schilke}},
  \bibinfo{author}{\bibfnamefont{C.}~\bibnamefont{Zimmermann}},
  \bibnamefont{and} \bibinfo{author}{\bibfnamefont{W.}~\bibnamefont{Guerin}},
  \bibinfo{journal}{Phys. Rev. A} \textbf{\bibinfo{volume}{86}},
  \bibinfo{pages}{023809} (\bibinfo{year}{2012}).

\bibitem[{\citenamefont{Aspelmeyer et~al.}(2010)\citenamefont{Aspelmeyer,
  Gr\"oblacher, Hammerer, and Kiesel}}]{aspelmeyer2010}
\bibinfo{author}{\bibfnamefont{M.}~\bibnamefont{Aspelmeyer}},
  \bibinfo{author}{\bibfnamefont{S.}~\bibnamefont{Gr\"oblacher}},
  \bibinfo{author}{\bibfnamefont{K.}~\bibnamefont{Hammerer}}, \bibnamefont{and}
  \bibinfo{author}{\bibfnamefont{N.}~\bibnamefont{Kiesel}},
  \bibinfo{journal}{J. Opt. Soc. Am. B} \textbf{\bibinfo{volume}{27}},
  \bibinfo{pages}{A189} (\bibinfo{year}{2010}).

\bibitem[{\citenamefont{Chan et~al.}(2011)\citenamefont{Chan, Mayer~Alegre,
  Safavi-Naeini, Hill, Krause, Gr\"oblacher, and Painter}}]{chan2011}
\bibinfo{author}{\bibfnamefont{J.}~\bibnamefont{Chan}},
  \bibinfo{author}{\bibfnamefont{T.~P.} \bibnamefont{Mayer~Alegre}},
  \bibinfo{author}{\bibfnamefont{A.~H.} \bibnamefont{Safavi-Naeini}},
  \bibinfo{author}{\bibfnamefont{J.~T.} \bibnamefont{Hill}},
  \bibinfo{author}{\bibfnamefont{A.}~\bibnamefont{Krause}},
  \bibinfo{author}{\bibfnamefont{M.}~\bibnamefont{Gr\"oblacher},
  \bibfnamefont{S.~ans~Aspelmeyer}}, \bibnamefont{and}
  \bibinfo{author}{\bibfnamefont{O.}~\bibnamefont{Painter}},
  \bibinfo{journal}{Nature} \textbf{\bibinfo{volume}{478}}, \bibinfo{pages}{89}
  (\bibinfo{year}{2011}).

\bibitem[{\citenamefont{Karrai et~al.}(2008)\citenamefont{Karrai, Favero, and
  Metzger}}]{karrai2008}
\bibinfo{author}{\bibfnamefont{K.}~\bibnamefont{Karrai}},
  \bibinfo{author}{\bibfnamefont{I.}~\bibnamefont{Favero}}, \bibnamefont{and}
  \bibinfo{author}{\bibfnamefont{C.}~\bibnamefont{Metzger}},
  \bibinfo{journal}{Phys. Rev. Lett.} \textbf{\bibinfo{volume}{100}},
  \bibinfo{pages}{240801} (\bibinfo{year}{2008}).

\end{thebibliography}
\end{document}